\documentclass[fleqn,12pt,twoside]{article}
\usepackage[headings]{espcrc1}

\usepackage{graphicx}
\usepackage[figuresright]{rotating}
\usepackage{psfig}
\usepackage{epsfig}
\usepackage{subfigure}

\title{Measurement of low mass dielectron continuum in $\sqrt{s_{NN}}$=200GeV Au-Au collisions with the PHENIX experiment at RHIC}

\author{Alberica Toia \address[suny]{Department of Physics and Astronomy, 
        State University of New York, Stony Brook, \\ 
        NY 11794-3800, USA}
for the PHENIX Collaboration}

\begin{document}
\maketitle

\begin{abstract}
The PHENIX experiment has performed the first measurement of the dielectron continuum in Au+Au collisions at RHIC energies.
Mass spectra are presented and compared with expectations from hadron decays.
\end{abstract}

\section{Introduction}
Electromagnetic probes are ideally suited to investigate hot and dense matter produced in high energy heavy ion collisions because they do not undergo strong interactions and thus probe the full time evolution of the collision. 
The dielectron continuum is rich in physics. It is expected that 
Dalitz decays of light hadrons, direct decays of vector mesons, as well as correlated charm decays contribute. 
At CERN SPS energies, the CERES experiment \cite{Cer} has discovered
an enhanced dielectron yield over the expected sources, interpreted as hint for in medium modification of hadron spectral functions, in particular of the $\rho$ meson.
Although $e^{+}e^{-}$ pairs are rare, the 0.24 nb$^{-1}$ collected by PHENIX in Au+Au collisions at $\sqrt{s_{NN}}$= 200 GeV in 2004 provide a significant sample to investigate the dielectron continuum. 
In the following, the analysis of 766M minimum bias events is presented.

\section{Di-Electron Analysis}
The PHENIX detector was designed for electron measurements. It combines excellent mass resolution (1\,\%) with powerful particle identification obtained by matching the reconstructed tracks with the information from a Ring Imaging Cherenkov detector (RICH) and an Electromagnetic Calorimeter (EMC). 
Electrons are identified by requiring at least 3 phototubes matched in the RICH and by correlating the energy $E$ measured in the EMC and the momentum $p$, parameterized in terms of $\frac{E-p}{p}$.
Pair cuts are applied to avoid sharing of detector hits:
shared hits in the drift chambers are removed with a cut on position parameters;
tracks which are parallel in the RICH 
are rejected by a cut on the angular difference. Whenever encountering such a pair, the complete event is rejected. \\
Pairs are created by combining all electrons and all positrons in one event. The overwhelming majority of these pairs is uncorrelated.
A statistical procedure is used to determine this
combinatorial background.
Since the PHENIX acceptance is different for like and unlike sign pairs, 
the combinatorial background is computed with a 
mixed event technique by pairing unlike sign tracks from different events with similar vertex position and centrality.
The shape of the background is reproduced with high precision by the mixed event technique, 
and tested with the ratio of the like sign spectra for real and mixed events which deviates from one by less than 0.1\,\%. \\
Four different methods have been tested to normalize the background
distributions.
The first normalizes the number of mixed events to the number of physical events. 
The second relies on the mean number of single electron and positron tracks which contribute to uncorrelated pairs
$\langle N_{+-} \rangle = \langle  n_+ \rangle \langle n_- \rangle$.
Under the assumption that electron and positron multiplicities are Poisson-distributed, and that the like sign
pairs are uncorrelated, one can derive a third normalization as
the geometrical mean of the like sign distributions
$\langle N_{+-} \rangle = 2\sqrt{N_{++}N_{--}}$.
Finally one can empirically normalize the background to the measured
like sign distributions and apply the same factor to the unlike sign distribution. 
All the normalizations agree within 0.5\,\%. 
The empiric normalization of the like sign pairs was chosen for the final results and 
a systematic uncertainty of 0.25\,\% was assigned. \\
After removal of the combinatorial background, the contribution of photon conversions was removed, cutting on
the orientation angle of the pair in the magnetic field.
Finally the spectra are corrected for efficiency such that the data represent the dielectron yield 
for which electron and positron are in the detector acceptance ($ | \eta | < 0.35$).

\section{Results}
\begin{figure}[htb]
 \begin{minipage}[l]{0.5\textwidth}
 \epsfig{file=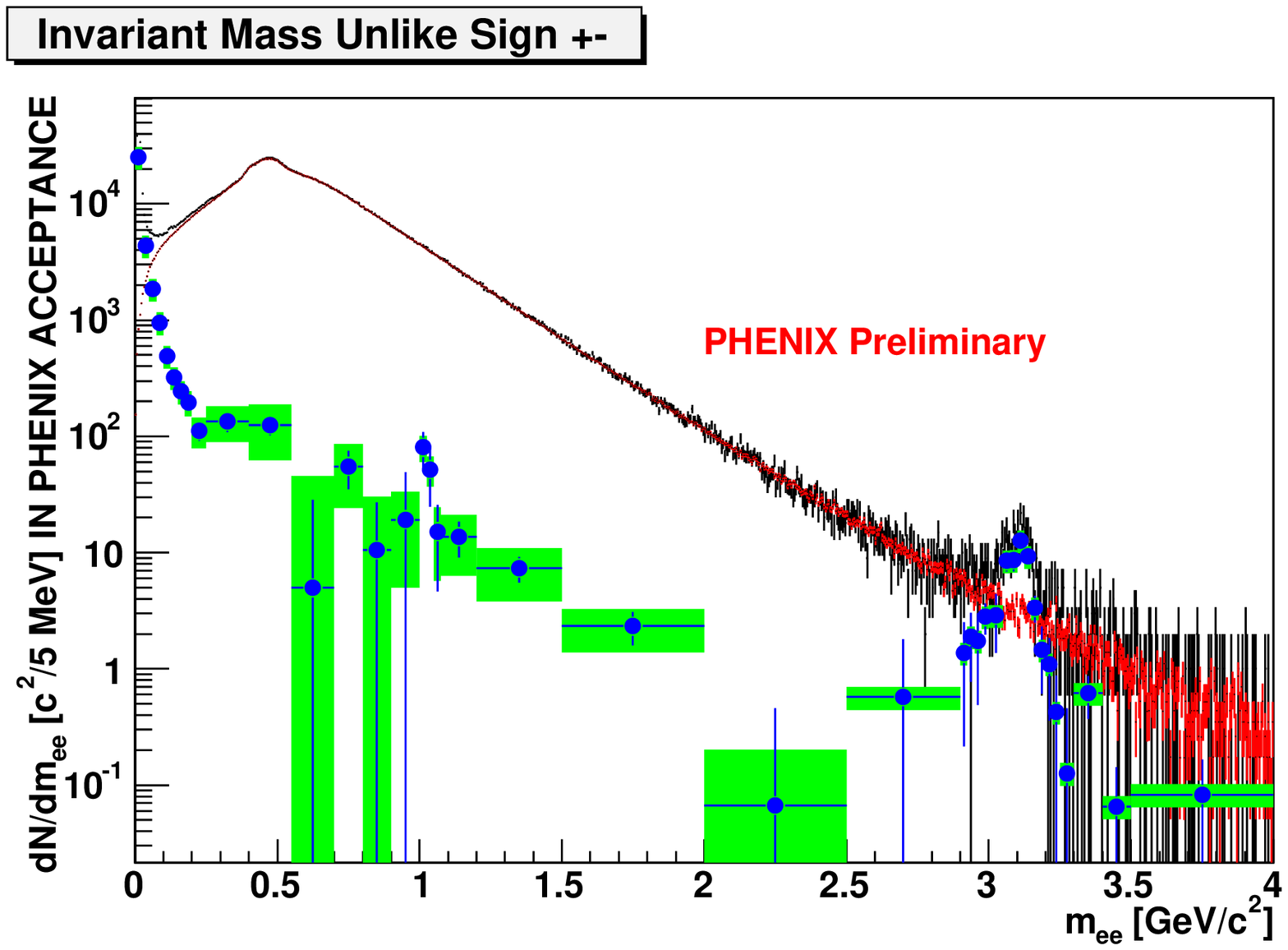,width=\textwidth}
 \end{minipage}
 \hfill
 \begin{minipage}[l]{0.5\textwidth}
 \epsfig{file=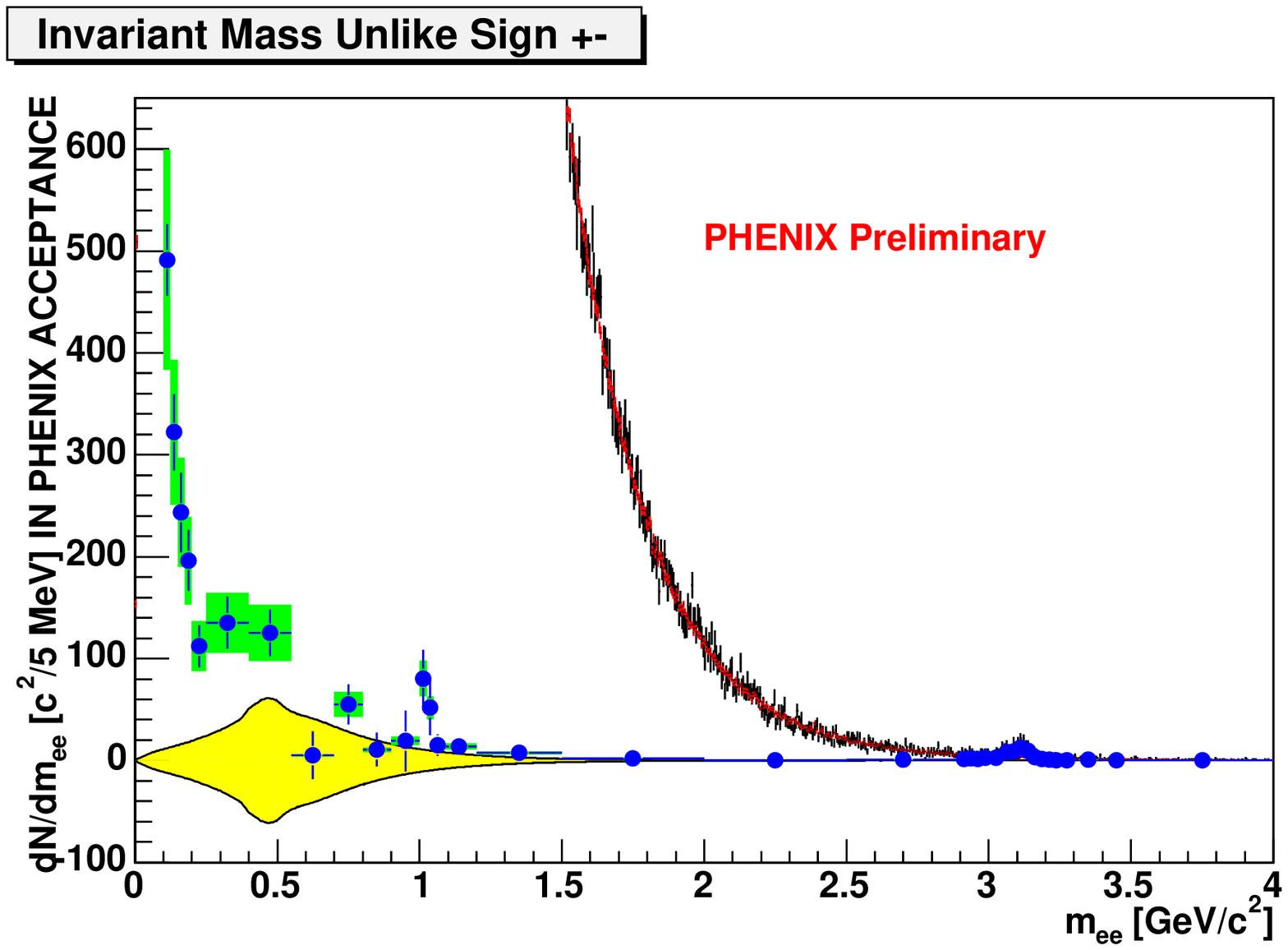,width=\textwidth}
 \end{minipage}
 \caption{Unlike sign mass spectrum: data (black), mixed events background (red), subtracted spectrum (blue), with systematic uncertainty. The left hand side shows the same on a linear scale, but the systematic uncertainty due to the background normalization is shown separately (colors in the online version).}
  \label{fig:mass}
\end{figure}

\begin{figure}[!b]
 \begin{minipage}[l]{0.5\textwidth}
 \epsfig{file=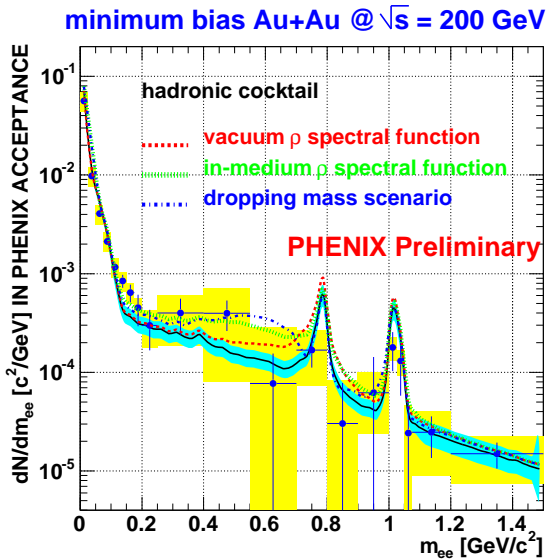, width=\textwidth}
 \caption{Data (systematic uncertainty in yellow) compared to the cocktail (systematic uncertainty in cyan) and theoretical predictions, where a $\rho$ spectral function is introduced, without (red) and with (blue and green) in-medium modifications (colors in the online version).
\label{fig:rapp}}
 \end{minipage}
 \hfill
 \parbox{0.5\textwidth}
{
\section{Conclusions}
The first measurement of the dielectron continuum at RHIC energy has been performed by the PHENIX experiment.
The dielectron mass spectrum is consistent both with the expectations from hadronic sources and with medium modified spectral functions. 
Due to the small signal to background ratio the systematic uncertainties are too large to make any conclusive statement. PHENIX is currently developing a new hadron blind detector system to address this issue\cite{Rav}.

}
\end{figure}

Figure \ref{fig:mass} shows the measured dielectron pairs, the background, and the subtracted spectra with unctertainty: the error bars in the left panel correspond to all the systematic errors, which are dominated by the background normalization.
The right panel shows separately the uncertainty arising from electron identification, efficiency, acceptance and run-by-run fluctuations, ($\approx$ 22\,\%) represented by the band around the data points, and the uncertainty on the background represented by the band around zero, which is the dominant source of the uncertainty due to the low signal to background ratio.\\
After background subtraction, a significant signal remains over the full invariant mass range, corresponding to an integral of $1.8 \cdot 10^5$ pairs, out of those 15,000 above the $\pi^0$ mass. 
\begin{table}[htb]
\caption{Measured and expected yield in different mass regions: low mass region ($0.15-0.7 \textrm{GeV}/\textrm{c}^2$) and intermediate mass region ($1.1-2.5 \textrm{GeV}/\textrm{c}^2$).}
\label{tab:yield}
\begin{tabular}{@{}lllll}
\hline
mass region ($\textrm{GeV}/\textrm{c}^2$) & Measurement ($10^{-5} \frac{counts}{event}$)  & Predictions ($10^{-5} \frac{counts}{event}$)  \\
\hline
0.15-0.7  &  $17.8 \pm 3.8 \pm 1.50$ & 12.3 \\
1.1-2.5   &  $0.67 \pm 0.50 \pm 0.11$ & 1.16 \\
\hline
\end{tabular}\\[2pt]
\end{table}
The expectations are based on a cocktail of hadron decay sources. 
The pion spectrum used as input is
determined by a parametrization of PHENIX charged and neutral
pion data. The spectra of the other mesons are determined from the pion
spectrum by $m_T$ scaling \cite{hep0508034}.
The systematic error, depending on the pion yield and the relative cross section of the other contributions, 
varies from 10\,\% to 25\,\%. \\
An additional source of dielectron pairs, which becomes dominant for invariant masses above 1\,$\textrm{GeV}/\textrm{c}^2$ is correlated charm production. 
\begin{figure}[htb]
 \begin{minipage}[l]{0.5\textwidth}
 \epsfig{file=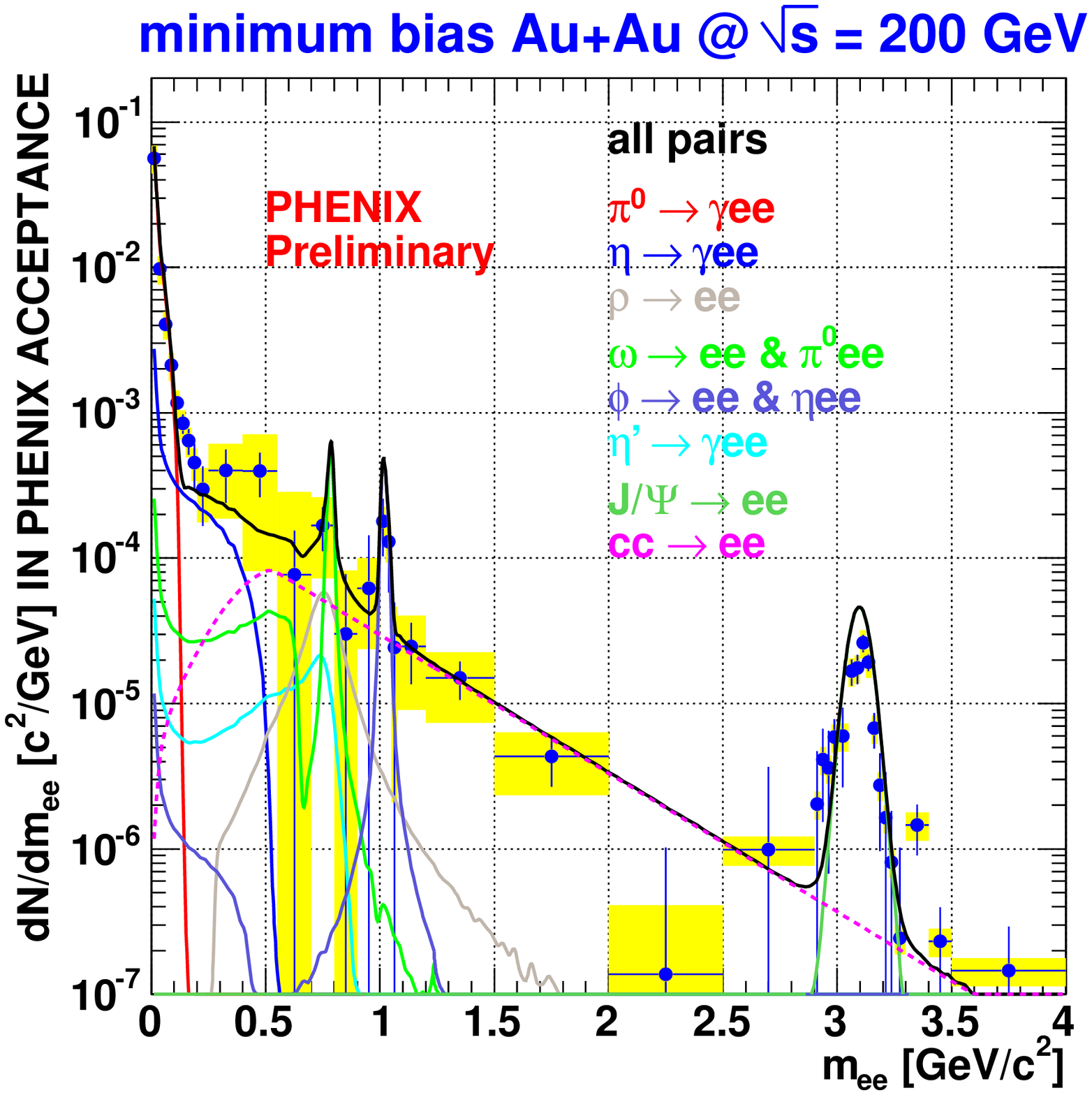, width=\textwidth}
 \end{minipage}
 \hfill
 \begin{minipage}[l]{0.5\textwidth}
 \epsfig{file=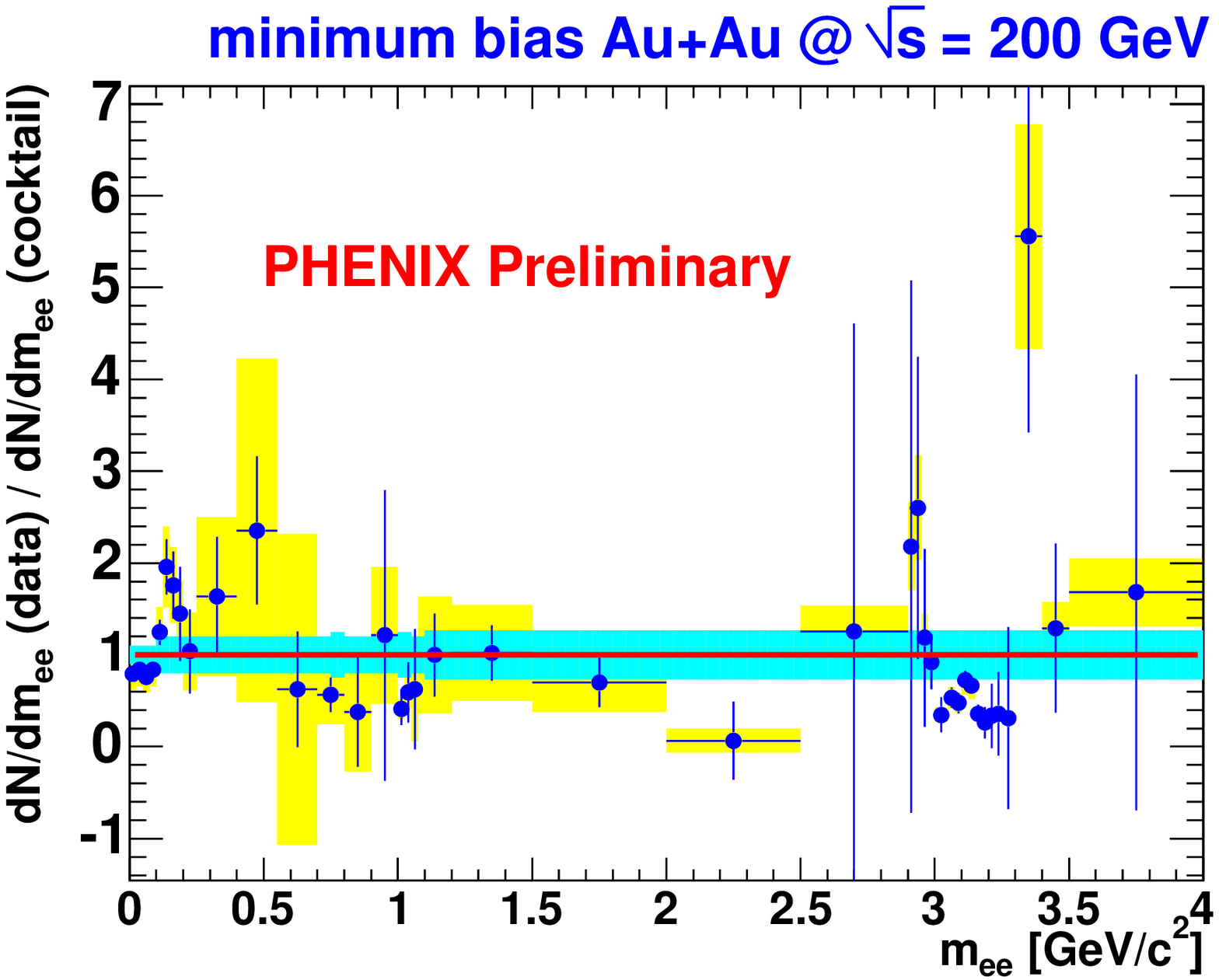, width=\textwidth}
 \parbox{\textwidth}
 {\caption{Data compared to a cocktail from hadronic sources and charm decays (left panel). Ratio of data to cocktail (right panel). Systematic uncertainties are shown in yellow (data) and cyan (cocktail) (colors in the online version). 
  \label{fig:cock}}}
 \end{minipage}
\end{figure}
This contribution has been simulated with PYTHIA, scaling the p+p equivalent $c\overline{c}$ cross section of 622$\pm$57$\pm$160 $\mu$barn to minimum bias Au+Au collisions proportional to the number of binary collisions ($258 \pm 25$) \cite{Prl94}. \\
The left panel of Figure \ref{fig:cock} shows the data compared to the cocktail.
The data are in good agreement with the cocktail over the full mass range. The $\omega$ and $\phi$ resonances are not fully reproduced, most likely because of the combined effect of an overstimate of the mass resolution in the event generator and low signal-to-background ratio. 
The data seem to lie above the cocktail in the region 0.3-0.8\,$\textrm{GeV}/\textrm{c}^2$ (see Table \ref{tab:yield} for a quantitative comparison). 
The systematic uncertainty however does not allow to draw any conclusive statement. 
The right panel shows the ratio of data to cocktail.\\
In Figure \ref{fig:rapp} the data are compared to theoretical predictions (\cite{rap1},\cite{rap2}), where the 
$e^+e^-$ invariant mass spectra $dN_{ee}/dM$ have been calculated 
using different in-medium $\rho$ spectral
 functions and an expanding thermal fireball model. 
Although the systematical uncertainty does not allow to state any deviation from the known sources, it is intriguing to observe the consistency of the data with predictions that include in-medium modifications in the same mass region where other experiments quantified such an effect.

\end{document}